# The role of impact and radiogenic heating in the early thermal evolution of Mars

S. Sahijpal and G. K. Bhatia

Department of Physics, Panjab University,

Chandigarh, India 160014

(sandeep@pu.ac.in)

**Abstract:**

The planetary differentiation models of Mars are proposed that take into account core-mantle and core-mantle-crust differentiation. The numerical simulations are presented for the early thermal evolution of Mars spanning up to the initial 25 million years (Ma) of the early solar system, probably for the first time, by taking into account the radiogenic heating due to the short-lived nuclides, $^{26}$Al and $^{60}$Fe. The influence of impact heating during the accretion of Mars is also incorporated in the simulations. The early accretion of Mars would necessitate a substantial role played by the short-lived nuclides in its heating. $^{26}$Al along with impact heating could have provided sufficient thermal energy to the entire body to substantially melt and trigger planetary scale differentiation. This is contrary to the thermal models based exclusively on the impact heating that could not produce widespread melting and planetary differentiation. The early onset of the accretion of Mars perhaps within the initial ~1.5 Ma in the early solar system could have resulted in substantial differentiation of Mars provide it accreted over the timescale of ~1 Ma. This seems to be consistent with the chronological records of the Martian meteorites.

# 1. Introduction

The solar system is considered to have formed around 4.568 billion years ago (Bouvier and Wadhwa 2010) as a result of the gravitational collapse of the protosolar molecular cloud. The gravitational collapse led to the formation of the sun that was surrounded by a solar nebula, a disk of gas and dust. The cooling of the nebular gas led to the condensation of grains. The coagulation and accumulation of these grains, followed by gravitational assisted accretion growth led to the formation of the planetesimals. The run-away growth of planetesimals within the initial few million years of the solar system formation is considered to have resulted in the formation of the planetary embryos like Mars (Murty 2014) with the dimensions less than the size of Earth. The gravitational accretion of the planetary embryos further resulted in the formation of terrestrial planets (e.g., Safronov 1969; Vityazev et al. 1978; Greenberg 1979, 1982; Kaula 1979; Coradini et al. 1983; Tonks and Melosh 1992; Weidenschilling 1997). A wide range of numerical studies have been performed to understand the accretion growth of the planetary embryos and terrestrial planets (Greenberg et al. 1978; Wetherill and Stewart 1989; Kokubo and Ida 1998; Kobayashi and Dauphas 2013).

Among all the terrestrial planets, the small size of Mars compared to Earth and Venus has been considered as one of riddles as a bigger planet is anticipated to form instead (Wetherill and Stewart 1989). The $^{182}$Hf-$^{182}$W isotopic systematics based on the SNC (Shergotty-Nakhla-Chassigny) meteorites suggests an early and rapid accretion of Mars preferably within the initial 4 million years (Ma) during the formation of the solar system (Kliene et al. 2004; Foley et al. 2005; Nimmo and Kliene 2007; Dauphas and Pourmand 2011). The rapid and early accretion indicates that Mars is probably a planetary embryo that escaped further substantial accretion to form an even bigger planet. Kobayashi and Dauphas (2013) have further suggested that Mars probably accreted rapidly in a massive disk from the accretion of small planetesimals with radii less than 10 km. The empirical support for the rapid and early accretion of Mars suggests an active role of the short-lived nuclides, $^{26}$Al and $^{60}$Fe, in causing wide-spread heating and planetary scale differentiation of Mars (Tang and Dauphas 2013, 2014). Based on the recent $^{60}$Fe-$^{60}$Ni systematic of five SNC meteorites and 11 chondrites a timescale of 1.9 (+1.7/-0.8) Ma has been inferred for the accretion and core formation of Mars during the early solar system, with a lower limit of 1.2 Ma (Tang and Dauphas 2014). It has been further proposed that Mars had probably acquired 44 % of its present size within the initial 1.8 Ma. Hence, at least the short-lived nuclide $^{26}$Al could have played a significant role in the early heating and probably planetary differentiation of the body. An attempt has been made in the present work to numerically simulate the planetary scale differentiation of Mars with the short-lived nuclides $^{26}$Al and $^{60}$Fe as the major heat sources along with the impact induced heating.

The widespread presence of the short-lived nuclide, $^{26}$Al ($\tau \sim$1 Ma; million years), in the early solar system has been based on the excess in its daughter nuclide, $^{26}$Mg, in Ca-Al-rich inclusions that are found in the primitive meteorites (MacPherson et al. 1995). Ca-Al-rich inclusions are considered to be the earliest condensed solar system phases in the solar nebula. An initial value of $5\times10^{-5}$ was deduced for $(^{26}\text{Al}/^{27}\text{Al})_{\text{ini.}}$ (the initial value of $^{26}$Al/$^{27}$Al) in the early solar system on the basis of analyses on Ca-Al-rich inclusions. Urey (1955) indicated the role of radionuclides present in the early solar system in causing the widespread heating, melting and even planetary differentiation of planetesimals and asteroids. There is substantial empirical evidence based on meteoritic studies that some of the planetesimals and asteroids underwent significant heating and planetary scale differentiation probably initiating within the initial few million years during the formation of the solar system (Chabot and Haack 2006; McCoy et al. 2006; Sahijpal et al. 2007). The discovery of yet another short-lived radionuclide, $^{60}$Fe ($\tau \sim$3.77 Ma) in the meteoritic phases (Tachibana and Huss 2003; Mostefaoui et al. 2005) further supported the role of short-lived nuclides in causing widespread thermal effects in planetesimals and asteroids. However, the inferred estimates of the $(^{60}\text{Fe}/^{56}\text{Fe})_{\text{ini.}}$ in the early solar system continue to be a debatable issue (see e.g., Tang and Dauphas 2014). A low value of $\sim 10^{-8}$ (Tang and Dauphas 2014) as well as a high value of $\sim 10^{-6}$ for $(^{60}\text{Fe}/^{56}\text{Fe})_{\text{ini.}}$ (Mostefaoui et al. 2005) have been proposed. The uncertainty in the $(^{60}\text{Fe}/^{56}\text{Fe})_{\text{ini.}}$ substantially influences the role of $^{60}$Fe as a planetary heat source.

A wide range of numerical models have been developed to understand the role of the two short-lived nuclides in causing wide range of thermal effects in the planetesimals and asteroids (Miyamoto 1991; Sahijpal et al. 1995; Sahijpal 1997; Merk et al. 2002; Hevey and Sanders 2006; Sahijpal et al. 2007; Gupta and Sahijpal 2010; Sahijpal and Gupta 2011; Elkins-Tanton et al. 2011; Moskovitz and Gaidos 2011; Henke et al. 2012, 2013; Neumann et al. 2011, 2012, 2014;

Sahijpal 2012; Šrámek et al. 2012; Weiss and Elkins-Tanton 2013; Sahijpal and Bhatia 2013). These thermal models cover a wide range of thermal processes within the planetesimals and asteroids that include thermal metamorphism (e.g., Miyamoto 1991; Sahijpal et al. 1995), the planetary scale differentiation (e.g., Sahijpal et al. 2007; Gupta and Sahijpal 2010) and aqueous alteration (Sahijpal and Gupta 2011; Sahijpal 2012). It should be noted that there is a general consistency among these various thermal models listed above in terms of the deduced thermal evolution except for the differences that originate due to the distinct choice of the various simulation parameters and the thermal evolution scenario(s). Merk et al. (2002) numerically solved the partial differential equation of heat transfer by incorporating linear accretion rate of planetesimals for the first time. Heavy and Sanders (2006) developed the thermal model for the differentiation of planetesimals assuming the accretion of planetesimals having an initial porosity of ~50%. They incorporated thermal convection subsequent to 50% silicate melting by raising the thermal diffusivity by three orders of magnitude. Sahijpal et al. (2007) numerically simulated the planetary differentiation of 20-270 km sized planetesimals. They studied two distinct scenarios for the differentiation of planetesimals. The growth of the metallic core and the extrusion of the $^{26}$Al-rich basaltic melt were studied for the first time for a linearly accreting planetesimal. Gupta and Sahijpal (2010) studied the differentiation of the asteroid 4 Vesta assuming a linear accretion rate. In this model, two different differentiation scenarios were studied. The one dealing with the formation of the basaltic achondrites in Vesta by 20 % partial melting of silicates, and the second scenario dealing with the formation of the basaltic achondrites from the residual melt left from the crystallization of the cooling magma ocean. The core-mantle differentiation was numerically executed in both scenarios. Sahijpal and Gupta (2011) performed numerical simulations to study the possibility of the formation of carbonaceous chondritic crust on partially differentiated asteroids. In this model, the piecewise accretion of planetesimals was considered, whereby, the initial accretion of the planetesimal occurs very rapidly followed by a slower accretion of the planetesimal. This result in an early formation of a sizable iron-core that is followed by the slow accretion of chondritic crust on the planetesimal surface. Elkins-Tanton et al. (2011) studied the thermal differentiation of planetesimals to understand the generation of core dynamo. Their results show that the bodies with radii larger than 100 km can have core dynamo lasting longer than 10 Ma.

Moskovitz and Gaidos (2011) studied the influence of the redistribution of $^{26}$Al due to the melt migration of basalt on thermal evolution. It was observed that for the melt viscosity less than 1 Pa s the extrusion of the melt towards the surface can occur to form a basaltic crust. Henke et al. (2012) studied the effect of initially porous body on the thermal evolution of planetesimal. The influence of compaction due to sintering which includes cold and hot pressing was considered. Neumann et al. (2012) studied the differentiation of planetesimal having melt fraction less than 50%. The segregation of iron melt from silicate melt was formulated using Darcy's law for the flow in porous media. In this model, two different liquidus temperatures of iron, i.e., 1233 K and 1700 K were considered to study its effect on the differentiation of planetesimals. The results show that for the liquidus temperature of 1233 K, the iron core formation takes places over a time less than 1 Ma which coincides with the results of Sahijpal et al. (2007) but for the scenario with the liquidus temperature of 1700 K, the core formation take much longer time that can go up to ~10 Ma. Sahijpal (2012) performed numerical simulations for the thermal evolution of planetesimals that accreted with substantial amount of ice in the early solar system. Distinct thermal evolutionary criteria were studied by varying the ice/dust mass fraction. Sahijpal and Bhatia (2013) studied the dependence of the growth rate of iron-core on the melt percolation velocity of the molten metallic blobs through the molten silicate. In this model, the velocity of melt in the range 0.0003-300 m yr$^{-1}$ was considered to understand the segregation of iron from silicate melt. For a melt percolation velocity < 0.01 m yr$^{-1}$, an extremely sluggish growth of iron-core was observed. The required velocity of > 0.01 m yr$^{-1}$ puts a stringent constraint for the formation of iron-core within the time anticipated from the chronological records of the differentiated meteorites. Neumann et al. (2014) have recently performed numerical simulations to study the formation of shallow magma ocean and crust on Vesta. Compared to the thermal models proposed by other workers, these models include several additional features such as advection, differentiation by porous flow and effective cooling of magma ocean. This work indicates the formation of a shallow magma ocean over which a basaltic crust forms. The core-mantle differentiation in this thermal models commence from the outside inward.

In the present work, we numerically simulate the early thermal evolution of Mars by taking into account the radiogenic heating due to the two short-lived nuclides, $^{26}$Al and $^{60}$Fe. This is probably the first attempt in this regard as the earlier thermal model for the early evolution of Mars (Senshu et al. 2002) considered only the impact induced heating and melting within the Mars during its accretion. The earliest epochs of the solar system is considered to have been dominated by heavy bombardments of the various planetary bodies by planetesimals and asteroids during their accretion stage. During the accretion, the impactors dumped large amount of energy and momentum to the planetary bodies that in turned provided thermal energy to cause widespread melting in large planetary embryos like Mars (Senshu et al. 2002) and terrestrial planets. However, in the case of accretion of asteroids and planetesimals, the impactors due to their small size could cause only localized heating (Keil et al. 1997). Nonetheless, the impactors could have played significant role in these cases by continuous reprocessing of the silicate crust. This aspect has been qualitatively studied in the thermal evolution of the minor planet, Vesta (Gupta and Sahijpal 2010) in terms of the anticipated rapid cooling of the magma ocean by continuous churning of the crust produced during the mantle-crust differentiation.

Senshu et al. (2002) adopted the approach developed by Tonks and Melosh (1992) to study the early thermal evolution of Mars during its accretion growth. Tonks and Melosh (1992) developed the formulation of diapir formation due to the giant impacts on the planet. A large impactor approaching with a very high velocity collides with the planet, a shock wave is produced which propagates spherically through both the bodies. The entire kinetic energy of the impactor is converted into heat. The thermal energy results in the segregation of metallic iron and silicate within the localized melt pond produced on account of heating. Tonks and Melosh (1992) pointed out that if the differential stress due to the high density of the metal body at the base of pond is more than the material threshold stress of the underlying mantle, the metallic iron melt will move towards the centre and form a metallic iron-core. The role of the radiogenic heating by $^{26}$Al and $^{60}$Fe in the thermal evolution of Mars was not included by Senshu et al. (2002). The thermal models developed by Senshu et al. (2002) were based exclusively on the impact induced heating and could not cause global planetary scale differentiation of Mars.

We make an attempt in the present work to numerically simulate the planetary differentiation of Mars on account of the impact induced heating during accretion (Senshu et al. 2002) and the radiogenic heating by the presence of the short-lived nuclides, $^{26}$Al and $^{60}$Fe (Sahijpal 1997; Sahijpal et al. 2007), at the time of early accretion of Mars in the solar system. As the previous models based exclusively on the impact induced heating could not explain the planetary scale differentiation of Mars, the present work makes an assessment on the role of the short-lived nuclides in causing planetary scale differentiation. We present here results of a representative set of numerical simulations that can be generalized to understand the detailed nature of the problem. Since, the recent $^{60}$Fe-$^{60}$Ni systematic of SNC meteorites indicate a timescale of 1.9 (+1.7/-0.8) Ma for the accretion and core formation of Mars during the early solar system (Tang and Dauphas 2014), the role of the short-lived nuclides become important to understand the early thermal evolution of Mars.

Several numerical simulations have been performed in the present work to study the dependence of the thermal models on some of the most critical parameters. These important parameters include, i) the onset time and the duration of the accretion of Mars in the early solar system, ii) the initial assumed abundance of $^{60}$Fe in the solar system, and iii) the melt-percolation velocity of the metallic iron blobs during their descent towards the center through the partially melted silicate during the segregation.

## 2. Methodology

The major framework involved in numerically simulating the early thermal evolution of Mars deals with solving the heat conduction partial differential equation with the radioactive heat sources, $^{26}$Al and $^{60}$Fe, using the finite difference method (Sahijpal et al 1995; Sahijpal 1997). Further, based on the formulation developed by Senshu et al (2002) we have incorporated the impact heating on account of accretion of Mars. The melting of the planetary body subsequently followed by planetary differentiation is numerically executed by adopting the approach based on the planetary differentiation of planetesimals and asteroids (Sahijpal et al. 2007; Gupta and Sahijpal 2010).

The partial differential equation associated with the heat conduction differential equation (1) involving the two radioactive heat sources was numerically solved using the finite difference method (Lapidus and Pinder 1982) with the classical explicit approximation (e.g., Sahijpal et al. 1995; Sahijpal 1997; Sahijpal et al. 2007; Gupta and Sahijpal 2010). The planet was assumed to be spherically symmetric. This reduces the dimensionality of the problem to one from three.

$$\frac{\partial T}{\partial t} = \kappa \frac{\partial^2 T}{\partial r^2} + Q_R \qquad (1)$$

Here, $Q_R$ is the heat produced due the decay of uniformly distributed short-lived radionuclides, $^{26}$Al and $^{60}$Fe, in an undifferentiated body. This contribution declines over time according to the mean-lives of the short-lived nuclides. $\kappa$ is the thermal diffusivity and 'T' is the temperature which is the function of space and time. The temperature dependence of the thermal diffusivity and specific heat was adopted for the un-melted body by following the criteria adopted by Sahijpal (1997) and Sahijpal et al. (2007). On contrary to the previous works on the thermal evolution of planetesimals and asteroids (e.g., Sahijpal et al. 2007), Mars was assumed to be sintered right from the onset time of its accretion. The porosity was assumed to be zero right from the onset of its accretion. This assumption was made in order to make comparison with the thermal models of Mars by Senshu et al. (2002) that assume sintered accretion of the body. This assumption could be justified as the impact induced heating will probably not only result in the sintering of the planetary body during accretion but would also cause melting.

The radionuclides, $^{26}$Al and $^{60}$Fe produce ~3.16 MeV (Ferguson 1958; Schramm et al. 1970; Sahijpal et al. 2007) and ~3 MeV (see e.g., Mostefaoui et al. 2005) energy per decay, respectively, that is available for heating the planetary bodies. The canonical initial value of $5\times10^{-5}$ for $(^{26}\text{Al}/^{27}\text{Al})_{\text{ini.}}$ (MacPherson et al. 1995) would correspond to an initial $^{26}$Al power generation per unit mass of undifferentiated planetary body to be $2.2\times10^{-7}$ W kg$^{-1}$ (Sahijpal et al. 2007), whereas, an initial value of ~$10^{-6}$ for $(^{60}\text{Fe}/^{56}\text{Fe})_{\text{ini.}}$ (see e.g., Mostefaoui et al. 2005) would correspond to an initial $^{60}$Fe power generation per unit mass of undifferentiated planetary body to be ~$2\times10^{-8}$ W kg$^{-1}$ (Sahijpal et al. 2007). Subsequent to differentiation of the planetary body the power generation rates will change on account of segregation of the radionuclides in core, mantle and crust. The estimated radiogenic heating is sufficient to cause widespread melting of a planetary body provided the body accretes within the initial couple of million years of the early solar system prior to the significant decay of the short-lived nuclides. It should be noted that the two short-lived nuclides were probably formed prior to the formation of solar system by a massive star (Sahijpal and Gupta 2009). As mentioned earlier, the inferred $(^{60}\text{Fe}/^{56}\text{Fe})_{\text{ini.}}$ in the early solar system is still uncertain (Mostefaoui et al. 2005; Tang and Dauphas 2014). A low value of ~$10^{-8}$ (Tang and Dauphas 2014) as well as a high value of ~$10^{-6}$ for the $(^{60}\text{Fe}/^{56}\text{Fe})_{\text{ini.}}$ (Mostefaoui et al. 2005) have been proposed. As the role of $^{60}$Fe as a planetary heat source would critically depend upon its initial estimated value, the lower adopted value would reduce any thermal contributions from $^{60}$Fe to effectively zero. Since, in the present work we focus only on the early thermal evolution of Mars up to the initial 25 Ma (million years) in the various simulations, the long lived radionuclide, $^{40}$K $^{235,238}$U, $^{232}$Th were not included as the heat sources due to their long mean-lives. These radionuclides contribute thermal energy gradually over an extended timescale of billion years rather than the initial tens of million years as considered in the present work. These radionuclides partition in mantle-crust subsequent to core-mantle-crust differentiation of a planetary body. The present estimated Martian crust radiogenic heating rate is ~5×10$^{-11}$ W kg$^{-1}$, confined mostly to the top ~50 km of the Martian surface (McLennan 2001; Hahn et al. 2011). At the time of the formation of Mars and its differentiation in the early solar system, this value extrapolates up to ~2×10$^{-10}$ W kg$^{-1}$ (Hahn et al. 2011). It should be noted that this value is almost identical to the initial heating rate estimates for a value of 1×10$^{-8}$ for ($^{60}$Fe/$^{56}$Fe)$_{ini.}$. As discussed in the following this low value of the initial estimates of $^{60}$Fe makes it an inefficient heat source for planetary processes over short interval of time ranging over few million years. However, subsequent to the core-mantle-crust differentiation of Mars in the early solar system, the long lived radionuclide, $^{40}$K $^{235,238}$U, $^{232}$Th played an important role as the heat sources during the last ~4.5 billion years thermal evolution (see e.g., Hahn et al. 2011; Tosi et al. 2013; Plesa et al. 2014).

The previous works performed earlier by our group were mostly focused on the thermal evolution of planetesimals and asteroids with a size range of 10-270 km in radii that is much smaller than the radius of 3400 km for Mars. The optimized choice of 0.3 km and one year was made for the spatial and temporal grid sizes, respectively, in the finite difference method in the case of smaller planetary bodies. However, in the present work in order to avoid large computational times for the Mars sized body, an optimized choices of 2 km and 10 years were made for the spatial and temporal grid sizes, respectively. We followed proper precautions to obtain stability and consistency in the numerical simulations with the present choices (Lapidus and Pinder 1982). As discussed earlier (Sahijpal et al. 2007), the details of the thermal evolution does not depend upon the choice of the spatial and temporal grid intervals unless there are stability issues in the numerical solutions.

The bulk composition of Mars was assumed to be identical to that of a H chondrite parent body (Jarosewich 1990; Sahijpal et al. 2007). The bulk oxygen, titanium, chromium and nickel isotopic compositions of the SNC meteorites indicate that Mars could have been made from H and EH chondrite planetesimals in the mass fraction of 55% and 45%, respectively (Tang and Dauphas 2014). However, the oxygen and nitrogen isotopic composition indicate that Mars probably accreted from H and EH chondrite planetesimals in the mass fraction of 26 % and 74 %, respectively (Mohapatra and Murty 2003). The choice of the H chondrite composition (Jarosewich 1990) in the present work is based on the fact that the melting and differentiation criteria for a body of this composition is well understood (Chabot and Haack 2006; McCoy et al. 2006). The bulk composition of Al and Fe was assumed to be 1.22 % and 27.8 %, respectively. The Al/Si and Fe(t)/Si ratios for the H chondrites are ~0.066 and 1.60, respectively (Jarosewich 1990). The metallic iron and the iron in FeS form constituted 16 % and 3 %, respectively, of the total bulk composition of the body that will eventually form the Fe-FeS metallic core of the body subsequent to differentiation. The remaining (8.8 %) iron was assumed as oxide in silicate that will eventually form the silicate-mantle. Subsequent to planetary differentiation, the entire un-decayed inventory of $^{26}$Al will eventually move to the silicate-mantle-crust, whereas, $^{60}$Fe will partition between the core and mantle according to the stable elemental iron abundance. The metallic and sulphide iron acquires an abundance of 86.36 % in the pure Fe-FeS metallic core, whereas, aluminium acquires an abundance of 1.56 % in the pure silicate mantle subsequent to core-mantle differentiation.

In order to study the early thermal evolution of Mars we assumed the planet to grow by accretion of planetesimals. As mentioned earlier, the dust grains formed in the solar nebula due to collisions eventually led to the formation of planetesimals. The further gravitational accretion of the planetesimals resulted in the formation of Mars-sized planetary embryos. In our model, we assumed a linear accretion law for the accretion of Mars from planetesimals (Merk et al. 2002, Sahijpal et al. 2007). The onset time ($T_{Onset}$) of the accretion of Mars was assumed to be in the range of 0.5-2 Ma (million years) subsequent to the formation of Ca-Al-rich inclusions that are considered to be the earliest condensed solar system phases in the solar nebula. The accretion onset was triggered by the accretion of planetesimals on a 2 km sized seed planetesimal. In the present work, Mars was assumed to accrete completely over a time span of 1 Ma in most of the simulations. This is consistent with the assumed accretion duration in the earlier work by Senshu et al. (2002) and the oligarchic growth of Mars (Weidenschilling 1997). We also performed a simulation with a comparatively slow accretion over a timescale of 2 Ma.

The accretion was numerically executed by systematically adding a spatial grid interval after a specific time interval to the existing spatial grid array in a linear manner. The initial surface temperature of the seed planetesimal at the onset of accretion was assumed to be 200 K (Senshu et al. 2002). Senshu et al. (2002) have performed detailed numerical simulations of the accretional growth of Mars, whereby the accreting planetesimals collide and impart impact kinetic energy to the accreting Mars at the surface. A power law distribution of the accreting planetesimals was assumed during the oligarchic growth, with an assumed minimum radius in the range of 30-80 km (Safronov, 1969) for the three set of simulations (Senshu et al. 2002). The average size of the planetesimals in these distributions is ~1.44 times the size of the planetesimal with the minimum size. As mentioned earlier, Senshu et al. (2002) adopted the criteria developed by Tonks and Melosh (1992) to study the conversion of the impactor kinetic energy into heat. A single impact results in the production of a pressurized isobaric core at the surface due to the transfer of the incident impactor velocity to the lattice particle velocity at the impact site. The generated shock wave raises the temperature within the isobaric core. This results in the shock induced heating, and probably melting of the compressed region within the isobaric core. The molten metallic blobs form at the base of these isobaric cores and commence their inward migration towards the center of Mars. The size of the isobaric core with respect to the impactor size was treated as a free parameter with a ratio in the range of 1-1.44. The accretion duration of the oligarchic growth of Mars was assumed to be 1 Ma in the simulations (Weidenschilling 1997). The impact induced temperature rise as a function of size of the growing Mars is presented in figure 6 of Senshu et al. (2002). In order to incorporate the impact induced heating of the accreting body during its accretion, we directly adopted this trend in the temperature rise on account of impact induced heating at the surface (see figure 6 of Senshu et al. 2002). This was numerically achieved during the accretion at each step by defining the surface temperature of the accreting body, i.e., protoMars, according to the temperature profile near the surface resulting from impact heating. As discussed in the next section, we are able to efficiently incorporate the parametric form of impact heating in our model during the accretion of Mars as anticipated by Senshu et al. (2002).

On contrary to the thermal models for the smaller bodies, e.g., planetesimals and asteroids where the impact heating is not effective (e.g., Sahijpal et al. 2007), the heating of the accreting Mars commenced not only from the center of the body on account of radiogenic heating but also from the surface of the accreting planet on account of impact heating. Hence, the melting could commence at the center as well as the outer regions of the accreting Mars. In the case of the thermal evolution of planetesimals and asteroids, the temperature range defining the solidus and the liquidus were assumed to be 1213-1233 K for Fe-FeS and 1450-1850 K for silicates (Sahijpal et al. 2007; Gupta and Sahijpal 2010). However, in the case of Mars that has comparatively higher internal pressures than asteroids, the melting temperatures of Fe-FeS and silicate were appropriately modified using the melting temperature dependence on pressure (Zahnle et al. 1988; Senshu et al. 2002) as given by the equation (2).

$$T_{melt} = T_0(1 + \frac{P}{P_0})^q \qquad (2)$$

where, $T_{melt}$ and $T_0$ are the melting temperature at pressures, P (Pa) and 0 Pa, respectively. $P_0$ and q are the constants with the values of $2\times10^{10}$ Pa and 0.36, respectively. During the accretion of Mars, the pressure was estimated at different spatial grid intervals and the melting temperatures were appropriately calculated according to the earlier formulation (Senshu et al. 2002). The latent heats of melting of Fe-FeS and silicates were assumed to be $2.7\times10^5$ J kg$^{-1}$ and $4.0\times10^5$ J kg$^{-1}$, respectively, and incorporated in the specific heat according to the criteria adopted in the earlier works (Merk et al. 2002; Sahijpal et al. 2007). The specific heat of the silicate melt and Fe-FeS melt was assumed to be 2000 J kg$^{-1}$ K$^{-1}$.

The molten magma ocean generated by large scale melting was assumed to acquire convection subsequent to 50 % melting of silicate. This was numerically achieved, partially, by considering an enhancement in the thermal diffusivity of the magma ocean by three orders of magnitude compared to the un-melted and consolidated body with zero porosity (Sahijpal et al. 2007; Gupta and Sahijpal 2010). The onset of convection will thus set an enhanced heat transfer. As mentioned in the earlier works (see e.g., Gupta and Sahijpal 2010) it is not possible to numerically simulate with more than three orders of magnitude enhancement in the thermal diffusivity as this results in numerical instabilities in the presently adopted finite difference method (Lapidus and Pinder 1982) with the classical explicit approximation. It has been recently pointed out that in order to efficiently incorporate convection the further enhancement would be essential (Neumann et al. 2014). This would reduce in the further heating of the interior of planetary body beyond the melting point corresponding to ~50% melting of silicate. Thus the thermal evolution of the planetary bodies will be significantly influenced subsequent to the onset of thermal convection. Since, in the present work we emphasis on the core-mantle differentiation that commence subsequent to ~40 % silicate melting, the thermal evolution of the planetary body till the core-mantle differentiation will not be significantly affected due to the adopted convective criteria. However, the thermal evolution over extended timescales dealing with the evolution of the magma oceans will be significantly influenced (Neumann et al., 2014). This is one of the reasons why we confine most of our discussions in the present work up to the initial ~25 Ma thermal evolution of Mars.

Subsequent to 40 % melting of silicate, the Fe-FeS melt blobs were assumed to sink in the silicate partial melt towards the center of the body as a diaper on account of density difference. The density of the bulk un-melted body was assumed to be 3500 kg m$^{-3}$, whereas, the density of the molten Fe-FeS blob was assumed to be 6700 kg m$^{-3}$. The segregation of the molten Fe-FeS from the partially melted silicate results in the onset of planetary scale differentiation of Mars. We adopted the formulation developed by Sahijpal et al. (2007) and Gupta and Sahijpal (2010) to execute the planetary scale differentiation. Based on the qualitative framework by Taylor et al. (1993), an average melt percolation velocity of the range of ~300 m yr.$^{-1}$ for the downward descend of the molten Fe-FeS was assumed in the planetary differentiation of planetesimals and asteroids. The growth rate of the Fe-FeS core depends upon this melt percolation velocity. Neumann et al. (2011, 2012, 2014) recently developed comprehensive numerical simulations for planetesimals and asteroids that take into account the Darcy's and Stroke's law for the melt percolation velocity of molten metal. It has been pointed out that the melt percolation velocities are essentially determined by Stoke's law that indicate almost instantaneous segregation (Neumann et al. (2014). Further, Sahijpal and Bhatia (2013) have recently developed thermal models that assume distinct averaged melt percolation velocities spanning over several orders of magnitude and imposed stringent constraints based on the chronological records of differentiated meteorites on the growth rate of the Fe-FeS cores in planetesimals and asteroids in case the short-lived nuclides were exclusively responsible for the planetary differentiation.

In the present work, the molten Fe-FeS blobs were moved towards the center with a specific melt percolation velocity that was considered as one of the simulation parameter. The molten Fe-FeS blobs generated within each spatial grid element of radial dimension of 2 km were moved towards the consecutive spatial grid towards the center. The downward descend of the molten metal pushes the partially melted silicate upward due to buoyancy. This eventually leads to the core-mantle segregation of Mars. As mentioned above, the rate of the core-mantle differentiation depends critically upon the settling velocity of the molten iron globules through the partially melted silicate (Neumann et al. 2014). We estimated the constraints on the settling velocity (v) by using the Stoke's law (equation 3).

$$v = (2/9) (\rho_g - \rho_m) g(r) a^2 / \eta \qquad (3)$$

where, $\rho_g$ and $\rho_m$ are the densities of molten metallic blob (6700 kg m$^{-3}$) and partially molten silicate (3500 kg m$^{-3}$), respectively. Here, g(r) is the acceleration due to gravity that varies from center to surface of the body. The surface value of g(r) was assumed to be 3.7 m s$^{-2}$. The corresponding value at the exterior of the central most spatial grid element that corresponds to the central 2 km was estimated to be ~0.002 m s$^{-2}$. It should be noted that there is a singularity at the center in terms of the acceleration due to gravity. The metallic blob that is moving towards the center can move to the exact center only due to its inertia. The gravity cannot move it to the exact central point of the body.

We deduced the silicate viscosity ($\eta$) above the melting point temperature corresponding to ~40% silicate melting in Mars using the standard procedure for viscosity determination (Giordano et al. 2008). The normalized silicate composition for viscosity determination was deduced by removing the metallic contents from the assumed H-chondrite composition (Jarosewich 1990) for Mars. The deduced value of the silicate viscosity can range from 1 - 0.1 Pa s. If we assume the typical radius (a) of the initial grains within the primitive chondritic body to be 0.1 mm (Taylor et al. 1993; Neumann et al. 2011) for the source of the generation of the initial molten metallic blobs, we deduce the value of the settling velocity of the molten iron through (~40 %) partially melted silicate to be 830 m yr.$^{-1}$ to 0.5 m yr.$^{-1}$ at the surface and the central 2 km of Mars, respectively, for the maximum deduced silicate viscosity of 1 Pa s. A reduction in the viscosity to 0.1 Pa s on account of further melting of silicate would increase the estimated settling velocities by an order of magnitude. It should be noted that an increase in the assumed initial grain size or the molten metallic blob size due to coalesce of different metallic blobs during their downward descent towards the center (Senshu et al. 2002) would result in a further increase in the settling velocity. The probability of coalesce increase as the molten metallic blobs move towards the center (Senshu et al. 2002). In general, we anticipate the settling velocities to be > 1 m yr.$^{-1}$ even towards the central regions of Mars, except for the central 2 km region. As mentioned earlier, the inertia of the molten metallic blob can alone take it towards the exact center of the body. In the central and outer regions of Mars the velocity goes around 100's m yr.$^{-1}$. In the present work, we ran several simulations to study the role of the melt percolation velocity on the core-mantle segregation. Senshu et al. (2002) estimated the settling velocity in the range of ~310 m yr.$^{-1}$ when the blob was passing through partially melted zone (Personal communication: Prof. H. Senshu). This means that most of metallic blobs settle at the base of partial melting zone within 300-500 years. However, at the base of the partial melting zone the descending speed rapidly decreased to ~0.003 m yr.$^{-1}$ or less because ambient silicate was highly viscous, resulting in a metallic layer in the thermal model (Senshu et al. 2002).

In majority of the simulations performed in the present work, we considered only the core-mantle differentiation without considering any possibility of crust formation. These models are referred in the present work as the CM (Core-Mantle) models (table 1). However, in one of the simulations, we initiated the crust formation subsequent to 20 % partial silicate melting. The core-mantle-crust differentiation was performed in this model, referred as CMC model (table 1). In the CMC model, the $^{26}$Al-rich basaltic melt was extruded upwards subsequent to 20 % partial silicate melting (Gupta and Sahijpal 2010). The upward extrusion of the $^{26}$Al-rich basaltic melt was performed through the entire planetary body over a timescale comparable to the mean life of $^{26}$Al (Taylor et al. 1993). We considered the radiogenic heating of the various regions of Mars during the upward descend of the basaltic melt.

As mentioned earlier we have concentrated specifically only up to the initial ~25 million years of the evolution of Mars. During the thermal evolution over this specific period, we have only included radiogenic heating due to short-lived nuclides and heat diffusive processes as these processes play significant role in the early thermal evolution. The convection in the molten iron and molten silicate was incorporated by raising the thermal diffusivity by three orders of magnitude during the ~50% melting of silicate. It should be noted that convection is a very fast mode of heat transfer. It is because of this fact that the temperature gradients across the body reduce significantly around ~50% silicate melting in all the simulations (Sahijpal et al. 2007; Gupta and Sahijpal 2010). It should be noted that in most of the simulations where we trigger only core-mantle differentiation (the CM models), the differentiation commence above ~40 % silicate melting that is almost around the value where the convection sets in. In most of these scenarios the exclusion of advection in our thermal models will not significantly influence the thermal evolution as the convection will bring in thermal equilibrium throughout the body over which the significant melting has taken place.

However, in the case of the thermal model with core-mantle-crust differentiation (the CMC model), the advection could become important during the extrusion of the $^{26}$Al-rich basaltic melt subsequent to ~20 % silicate melting. We ran one simulation to understand the role of $^{26}$Al basaltic melt extrusion on the thermal evolution. We could incorporate the radiogenic heating of the various planetary regions by the passage of the $^{26}$Al-rich basaltic melt.

## 3. Results & Discussion

We present results of the numerical simulations of the early thermal evolution of Mars by incorporating the influence of radiogenic heating due to the short-lived nuclides along with the impact induced melting. The early thermal evolution of Mars results in its planetary scale differentiation with the formation of an iron-core and a silicate-mantle. We have performed a parametric study to understand the dependence of the extent of planetary differentiation of Mars on some of the most important parameters that include the onset time and duration of the accretion of Mars in the early solar system, the initial abundance of $^{60}$Fe in the solar system due to its uncertain value and the melt-percolation velocity of the metallic iron blob through silicate melt during its descent towards the center of the body during the segregation (table 1). The various simulations performed in the present work are named according to the most critical simulation parameters in order to facilitate a quick reference (table 1). The nomenclature is defined as; [Model type]-[$T_{Onset}$]-[$T_{acc}$]-[($^{60}$Fe/$^{56}$Fe)$_{ini.}$, in units of $10^{-7}$]-[Assumed average velocity of metallic blobs during their descent towards the center of Mars]. The model type CM refers to the scenario where only the core-mantle segregation was performed. In the CMC model the $^{26}$Al-rich basaltic melt extrusion towards the outer region was incorporated subsequent to 20

% silicate melting. This could result in subsequent crust formation. The interiors of the body get depleted in $^{26}$Al, and are thus further deprived of its heating. However, as the $^{26}$Al-rich pockets move upwards through the planetary body over the assumed timescale of ~1 Ma, the different regions are heated up.

The first two parameter dealing with the onset time and the duration of the accretion of Mars are extremely critical parameters as these parameters determine the abundance of the short-lived nuclides, $^{26}$Al and $^{60}$Fe that survived against their radioactive decay to play a major role in heating the planetary body. It has been shown earlier that the onset time of the accretion of a planetary body beyond 3 Ma significantly reduces the potency of the short-lived nuclides to cause widespread heating and melting of the planetesimals and asteroids (Sahijpal et al. 2007). It should be mentioned here that the melting points within the larger bodies like Mars are raised on account of high pressures prevailing within the bodies compared to the planetesimals and asteroids where the internal pressures does not alter the melting points of metals and silicates. This would reduce the timescale for the effectiveness of the short-lived nuclides in causing widespread melting and differentiation in Mars compared to planetesimals and asteroids. In the present work, we ran simulations with an onset time of accretion in the range of 0.5-2 Ma. An earlier accretion than this would result in an earlier and more extensive wide-range melting of the planetary body. The accretion duration was assumed to be 1 Ma in most of the simulations except for the simulation CM-1-2-0.1-200, where we assumed an accretion duration of 2 Ma (table 1).

The initial abundance of $^{60}$Fe is also one of the critical parameters, as its inferred value in the early solar system continues to be a matter of debate (Tachibana and Huss 2003; Mostefaoui et al. 2005; Sahijpal et al. 2007; Tang and Dauphas 2014). We assumed an initial value of $10^{-8}$ for $(^{60}Fe/^{56}Fe)_{ini.}$ in most of the simulations at the time of condensation of Ca-Al-rich inclusions to be consistent with the recent work by Tang and Dauphas (2014). However, we performed a simulation CM-2-1-10-200 with a higher value of $10^{-6}$ for $(^{60}Fe/^{56}Fe)_{ini.}$ (table 1). An initial value of $5\times10^{-5}$ was assumed for $(^{26}Al/^{27}Al)_{ini.}$ in the early solar system for all the simulations at the time of formation of the earliest solar system phases, viz., the Ca-Al-rich inclusions (MacPherson et al. 1995). The time associated with the condensation of Ca-Al-rich inclusions marks the beginning of the temporal scale for all the processes related with the solar system evolution.

As mentioned in the earlier section, the melt percolation velocity of the molten metal during its descent towards the center of the planetary body through the partially melted silicate determines the core-mantle segregation rate. We assumed the melt percolation velocity as a simulation parameter. We performed a set of four simulations, CM-1.3-1-0.1-200, CM-1.3-1-0.1-2, CM-1.3-1-0.1-0.4 and CM-1.3-1-0.1-0.1, with a wide range of an assumed average melt percolation velocity to understand its dependence on the thermal evolution of Mars up to the initial 25 Ma (table 1).

The results obtained from the simulations of the thermal evolution of Mars are presented in figures 1-7. Some of the thermal profiles during the accretion and evolution of Mars up to the initial ~25 Ma of the formation of the solar system are presented along with the associated temporal evolution of the segregation of iron. The initial total iron content (27.8%) of the un-melted Mars subsequent to melting and onset of planetary differentiation is segregated into metallic [Fe (16%) + FeS(3%)] and oxide [Fe (8.8%)] fractions. While the molten metallic fraction initiates its downward descent towards the planetary center, the partially melted silicate fraction is either left behind or move upwards due to its buoyancy during the segregation. On the basis of the argument made in the previous section we assumed an average melt percolation velocity of 200 m yr.$^{-1}$ for the movement of the molten metal towards the planetary center in all the results presented in figures 1-7. This would eventually lead to the formation of an iron core with 86.36 % iron and the remaining content as sulphur. Further, on the basis of a set of four simulations with distinct melt percolation velocities (table 1, figure 8), we make an argument that for the choice of any assumed average velocity > 1 m yr.$^{-1}$ the growth rate of the iron-core formation remains unaltered.

The simulation CM-0.5-1-0.1-200, with the onset time of accretion (T$_{Onset}$) of 0.5 Ma and an ($^{60}$Fe/$^{56}$Fe)$_{initial}$ value of 10$^{-8}$ (figure 1) results in a complete core-mantle differentiation of Mars within the initial 1.66 Ma. As mentioned in the previous section due to the onset of heating simultaneously at the center because of the short-lived nuclides (Sahijpal 1997; Sahijpal el al. 2007) and in the outer regions because of impacts (Senshu et al. 2002), the melting of the body commence in the two regions. This eventually triggers metal-silicate segregation in the corresponding regions. The metal-silicate segregation noticed around 1.31 Ma (figure 1) forms a metallic shell identical to the one noticed earlier by Senshu et al. (2002). This metallic shell grows over time and moves towards the center and eventually merges around 1.66 Ma with the ongoing core-mantle segregation occurring at the center. In the case of the thermal model proposed by Senshu et al. (2002), the downward movement of the metallic shell was arrested due to the lower temperatures prevailing towards the center on account of the absence of the radioactive heating. However, in the present work, we have successfully achieved the complete core-mantle segregation. These models along with the empirical temporal constraints based on the Martian meteorites can explain the planetary scale differentiation of Mars in the early solar system (Kliene et al. 2004; Foley et al. 2005; Nimmo and Kliene 2007; Dauphas and Pourmand 2011; Tang and Dauphas 2013, 2014). Since, the $(^{60}Fe/^{56}Fe)_{initial}$ value of $10^{-8}$ is low to cause any efficient heating, the subsequent heating of the metallic core after 1.66 Ma is almost over. However, the silicate mantle continues to heat due to the presence of $^{26}Al$ in the mantle (figure 1).

In comparison to the differentiation scenario CM-0.5-1-0.1-200 (figure 1), the delay in the onset time of accretion of Mars by 0.5 Ma for the model CM-1-1-0.1-200 (figure 2) would result in a delay in the appearance of the metallic concentric ring by at least 0.5 Ma and the merger of the metallic shell with the metallic core by at least 1.31 Ma. A further delay of 0.3 Ma in the onset time of the accretion for the model CM-1.3-1-0.1-200 (figure 4) will result in a delay in the appearance of the metallic concentric ring by at least 0.8 Ma with respect to the model CM-0.5-1-0.1-200 (figure 1). The merger of the metallic shell with the metallic core occurs around 6.4 Ma (figure 4) in the model CM-1.3-1-0.1-200.

In the case of the model CM-2-1-0.1-200 (figure 5) with the onset time of accretion ($T_{Onset}$) of 2 Ma and a low $(^{60}Fe/^{56}Fe)_{initial}$ value of $10^{-8}$ (Tang and Dauphas 2014) the formation of the metallic core does not occur due to the insufficient short-lived radioactive heating. This scenario is almost identical to the thermal model proposed by Senshu et al. (2002) as it results in the formation of a metallic shell rather than a core. The inward progress of the metallic shell is halted due to the inadequate silicate melt fraction below the threshold level of ~40 %. Thus, in a way we have also reproduced the thermal model proposed by Senshu et al. (2002) with a melt percolation velocity of 200 m yr.$^{-1}$ as mentioned in the previous section.

Even though the scenario with an onset time of 2 Ma for the accretion of Mars with an $(^{60}Fe/^{56}Fe)_{initial}$ value of $10^{-8}$ (Tang and Dauphas 2014) (figure 5) does not result in the formation of metallic core, an increased $(^{60}Fe/^{56}Fe)_{initial}$ value of $10^{-6}$ (Mostefaoui et al. 2005) in the model CM-2-1-10-200 could however result in large scale planetary differentiation (figure 6). Thus, we infer that the accretion of Mars around 2 Ma in the early solar system could have resulted in large scale planetary differentiation provided the initial value of $^{60}Fe$ was high (Mostefaoui et al. 2005). However, the lower value of $10^{-8}$ for $(^{60}Fe/^{56}Fe)_{initial}$ (Tang and Dauphas 2014) would set the onset time limit to ~1.3 Ma in the early solar system. In any case, we rule out the possibility of differentiation beyond an onset time of accretion of 3 Ma exclusively due to the short-lived nuclides and impact heating.

A delayed accretion of Mars also results in a sluggish growth of the iron core as is evident from the simulation CM-1-2-0.1-200 (figure 3). Compared to the model CM-1-1-0.1-200 (figure 2) with the accretion timescale of 1 Ma, the model CM-1-2-0.1-200 (figure 3) with the accretion duration of 2 Ma, exhibits a retarded growth of the iron core. The merger of the molten metallic shell and the core does not commence in the latter case even around 25 Ma due to the insufficient partial melting of silicate. This imposes stringent constraints on the role of short-lived nuclides in case the accretion of Mars occurs beyond a timeframe > 1 Ma. However, it should be noted that the oligarchic growth of Mars has been suggested to occur over a timescale up to 1 Ma (Weidenschilling 1997)

We tried a single simulation CMC-0.5-1-0.1-200 (figure 7) with the possibility of the formation of crust. Subsequent to the 20% partial melting of silicate, the $^{26}Al$-rich basaltic melt was generated and moved towards the outer regions over an assumed timescale comparable to the mean life of $^{26}Al$ (Taylor et al. 1993). The central early depletion of $^{26}Al$ results in a reduction in the rate of temperature rise at the center. However, the extrusion of the basaltic melts in the outer regions results in an assisted temperature rise in those regions (figure 7). It should be noted that the growth of the iron core in this scenario occurs not exactly from the center as has been noticed recently by Neumann et al. (2014). This could be a promising model to understand the early thermal evolution of Mars if we incorporate exact constraints on the extrusion velocity of the $^{26}Al$-rich basaltic melt, advection and a proper treatment of the mantle and crust early evolution.

A set of four simulations were run with distinct set of melt percolation velocity of the molten iron during its descent towards the center of the planetary body to study its dependence upon the core-mantle segregation rate (figure 8). These assumed average velocities span over three orders of magnitude. It should be noted that an almost identical trends are observed in the case of velocities of 200 m yr.$^{-1}$ and 2 m yr.$^{-1}$ and to an extent even in the case of 0.4 m yr.$^{-1}$ (figure 8). Thus, instead of the use of an assumed average velocity in our simulations if the velocity is varied over this range due to variation in the acceleration due to gravity, the results do not differ significantly. The rate of core-mantle differentiation appears to be almost independent of the choice of velocity. Thus, the growth rate of the iron-core formation will explicitly depend upon the rate at which the partial melting of silicate takes place in the body in the central regions. This in turn depends upon the accretion time and duration of Mars along with the ($^{60}$Fe/$^{56}$Fe)$_{ini.}$ value.

As discussed elaborately in the previous section, we do not anticipate melt percolation velocities to be less than ~1 m yr.$^{-1}$ through a ≥40% partially melted silicate mush even close to the center of the body where the gravitational field is weak. However, the further reduction in the velocity to a value ~ 0.1 m yr.$^{-1}$ significantly reduces the possibility of even planetary scale differentiation of Mars (figure 8). This is not anticipated even with the differentiation scenario proposed by Senshu et al. (2002) as long as the molten metallic blobs can efficiently move through the ≥40% partially melted silicate mush. During their descent through the silicate mush, the molten metallic blobs will coalesce with other blobs to form bigger blobs. The probability of coalescence increases with depth (Senshu et al. 2002). The bigger blobs will continue to move with larger velocities regardless of the reduction in the gravity. Thus, we do not anticipate a substantial reduction in the melt percolation velocity towards the central region.

## 4. Conclusions

The planetary differentiation of Mars has been numerically simulated with the incorporation of the radioactive heating due to the short-lived nuclides $^{26}$Al and $^{60}$Fe and impact induced heating during the accretion of Mars in the early solar system. On contrary to the earlier thermal models, we could demonstrate the planetary scale differentiation of Mars, provided it accreted rapidly in the early solar system. The physical segregation of molten metal and partially melted silicate was numerically executed for two distinct set of models related with the core-mantle and core-mantle-crust differentiation. It was observed that the onset of the accretion of Mars should commence within the initial 1.5 Ma (million years) of the formation of the solar system in order to cause early differentiation of Mars as is indicated by chronological records of the Martian meteorites. Further, in order to cause substantial differentiation, the accretion of Mars should occur over a duration of ~1 Ma that is expected from the oligarchic growth of Mars.

**Acknowledgements:** We are extremely grateful to the reviewers for constructive criticism and making several comments and suggestions that led to significant improvement of the manuscript. We are thankful to Prof. H. Senshu for useful discussion regarding their estimates of the melt percolation velocity of the molten metallic blobs. This work is supported by the PLANEX (ISRO) research project.

Table 1. *The details of the thermal models along with the choice of parameters.*

| S. No. | Model$^\$$ | $T_{Onset}$# (Ma) | $T_{acc.}$* (Ma) | $(^{60}Fe/^{56}Fe)_{ini.}$ $^£$ | Velocity$^\S$ (m/yr.) |
|---|---|---|---|---|---|
| 1 | CM-0.5-1-0.1-200 | 0.5 | 1 | $10^{-8}$ | 200 |
| 2 | CM-1-1-0.1-200 | 1 | 1 | $10^{-8}$ | 200 |
| 3 | CM-1-2-0.1-200 | 1 | 2 | $10^{-8}$ | 200 |
| 4 | CM-1.3-1-0.1-200 | 1.3 | 1 | $10^{-8}$ | 200 |
| 5 | CM-1.3-1-0.1-2 | 1.3 | 1 | $10^{-8}$ | 2 |
| 6 | CM-1.3-1-0.1-0.4 | 1.3 | 1 | $10^{-8}$ | 0.4 |
| 7 | CM-1.3-1-0.1-0.1 | 1.3 | 1 | $10^{-8}$ | 0.1 |
| 8 | CM-2-1-0.1-200 | 2 | 1 | $10^{-8}$ | 200 |
| 9 | CM-2-1-10-200 | 2 | 1 | $10^{-6}$ | 200 |
| 10 | CMC-0.5-1-0.1-200 | 0.5 | 1 | $10^{-8}$ | 200 |

$ The nomenclature of the thermal model is read as;

[Model type]-[$T_{Onset}$]-[$T_{acc}$]-[$(^{60}Fe/^{56}Fe)_{ini.}$, in units of $10^{-7}$]-[Assumed velocity of metallic blobs during their descent towards the center of Mars]. The CM model refers to core-mantle segregation model. In the CMC model the $^{26}$Al-rich basaltic melt extrusion towards the outer region was incorporated subsequent to 20 % silicate melting. This could result in crust formation. The majority of the simulation parameters other the one discussed in the table are almost fixed according to the adopted thermodynamical criteria (see e.g., Sahijpal et al. 2007; Gupta and Sahijpal 2010).

# $T_{Onset}$ (Ma) is the assumed onset time of the accretion of Mars in the early solar system subsequent to the formation of Ca-Al-rich inclusion. The $^{60}$Fe-$^{60}$Ni systematic of SNC meteorites suggests a timescale of 1.9 (+1.7/-0.8) Ma for the accretion and core formation of Mars during the early solar system (Tang and Dauphas 2014).

* $T_{acc.}$ (Ma) is the assumed accretion duration of Mars. The oligarchic growth of Mars suggest a time duration up to 1 Ma (Weidenschilling 1997).

£ The assumed $(^{60}Fe/^{56}Fe)_{ini.}$ value at the time of formation of Ca-Al-rich inclusions. This can range from ~$10^{-8}$ (Tang and Dauphas 2014) to ~$10^{-6}$ (Mostefaoui et al. 2005). A value of $5\times10^{-5}$ was assumed for the $(^{26}Al/^{27}Al)_{ini.}$ at the time of formation of Ca-Al-rich inclusions (MacPherson et al. 1995).

§ The assumed average melt percolation velocity of the molten metallic blobs through partially melted silicate (equation 3). A typical value of 200 m yr.$^{-1}$ was assumed in most of the simulations.

**Figure captions:**

Figure 1. a) A selected set of thermal profiles for the model CM-0.5-1-0.1-200 of Mars during the initial stages of the formation of the solar system. The onset time of accretion ($T_{Onset}$) of 0.5 Ma and an accretion duration of 1 Ma were assumed. A $(^{60}Fe/^{56}Fe)_{initial}$ value of $10^{-8}$ was assumed at the time of condensation of Ca-Al-rich inclusions, the earliest solar system grains, that marks the beginning of the temporal scale. All timescales are measured with respect to the condensation of Ca-Al-rich inclusions, the earliest solar system grains. b) The associated temporal evolution of the segregation of the initial total iron content (27.8 %) of the un-melted Mars into metallic [Fe (16%) + FeS(3%)] and oxide [Fe(8.8%)] subsequent to melting. The molten metal moves towards the center with an assumed average melt percolation velocity of 200 m yr.$^{-1}$. This would eventually lead to the formation of an iron core with 86.36 % iron and the remaining content as sulphur. The $^{26}$Al-rich basaltic melt extrusion towards the outer region was not incorporated.

Figure 2. Identical to fig. 1, except with $T_{Onset}$ of 1.0 Ma for the thermal model CM-1-1-0.1-200.

Figure 3. Identical to fig. 1, except with accretion duration of 2.0 Ma for the thermal model CM-1-2-0.1-200.

Figure 4. Identical to fig. 1, except with $T_{Onset}$ of 1.3 Ma for the thermal model CM-1.3-1-0.1-200.

Figure 5. Identical to fig. 1, except with $T_{Onset}$ of 2.0 Ma for the thermal model CM-2-1-0.1-200.

Figure 6. Identical to fig. 5, except with a $(^{60}Fe/^{56}Fe)_{initial}$ value of $10^{-6}$ for the thermal model CM-2-1-10-200.

Figure 7. Identical to fig. 1, except with the $^{26}$Al-rich basaltic melt extrusion towards the outer region in the thermal model CMC-0.5-1-0.1-200.

Figure 8. The temporal evolution of the segregation of the initial total iron content of the un-melted Mars into metallic [Fe+FeS] and oxide [Fe] subsequent to melting. A set of fours simulations with distinct choices of the assumed average melt percolation velocity spanning over four orders of magnitude were run. A $(^{60}Fe/^{56}Fe)_{initial}$ value of $10^{-8}$ and $T_{Onset}$ of 1.3 Ma were assumed in all cases, identical to the CM-1.3-1-0.1-200 (figure 4).

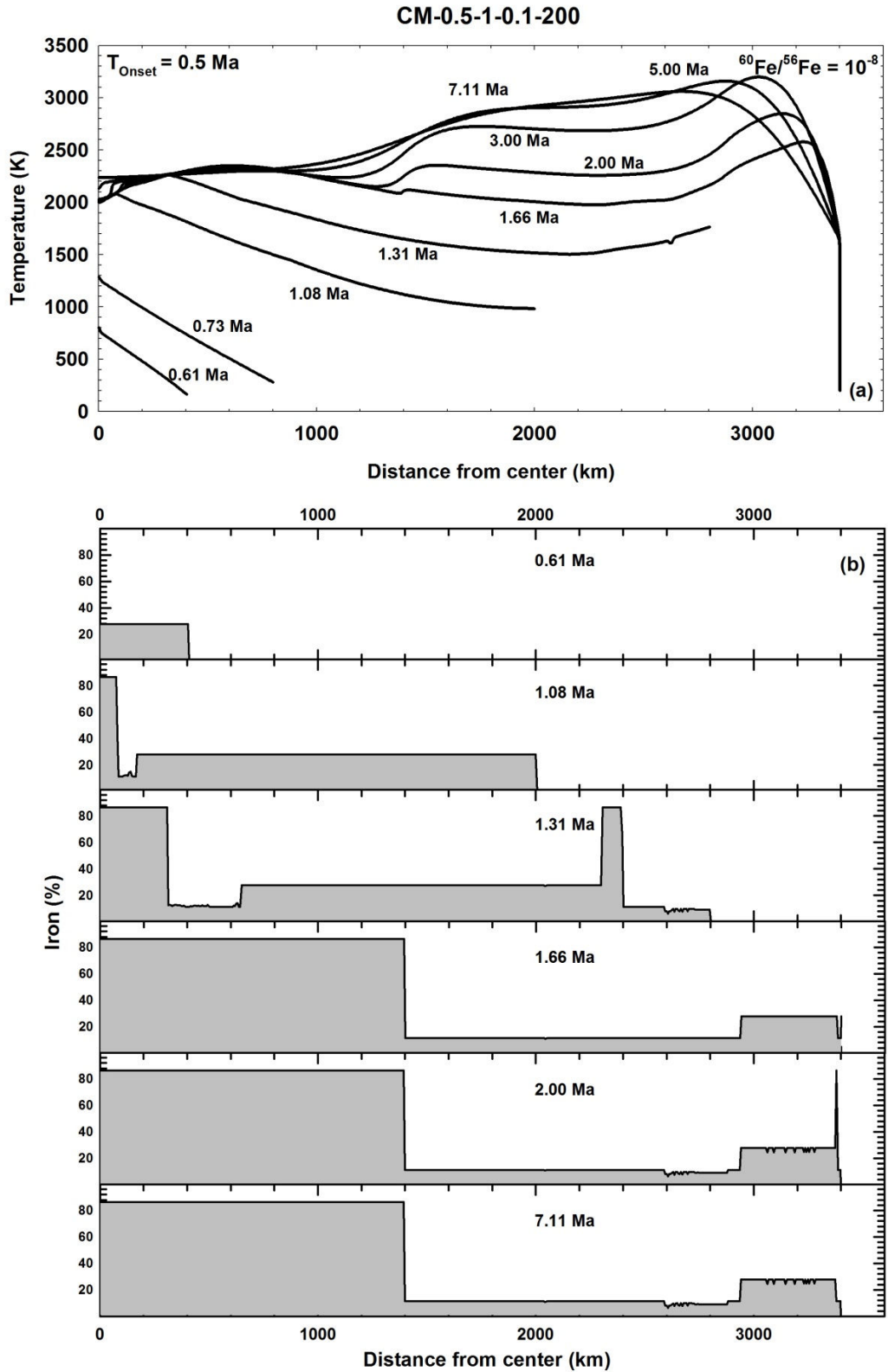

**Fig. 1.**

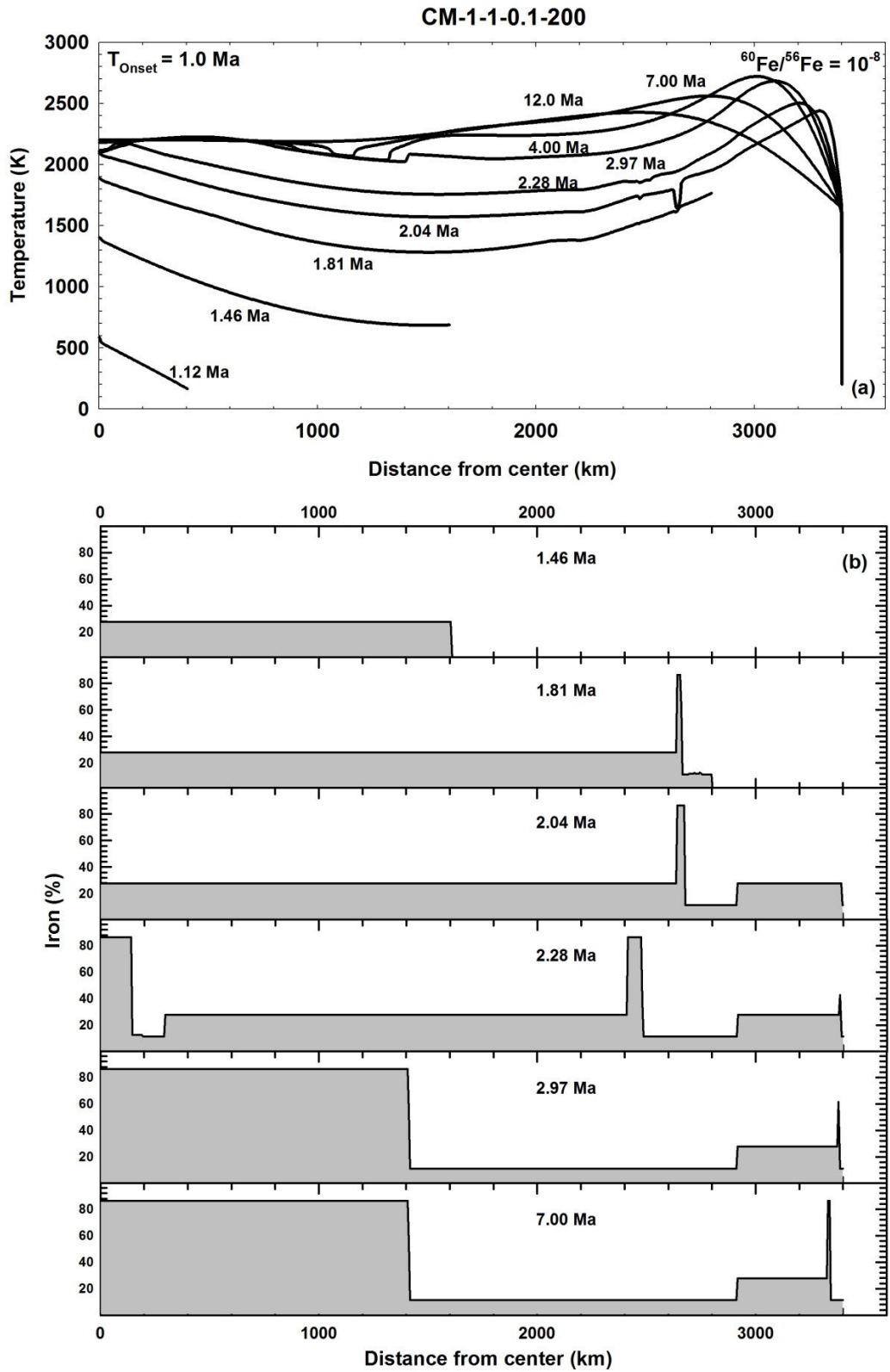

**Fig. 2.**

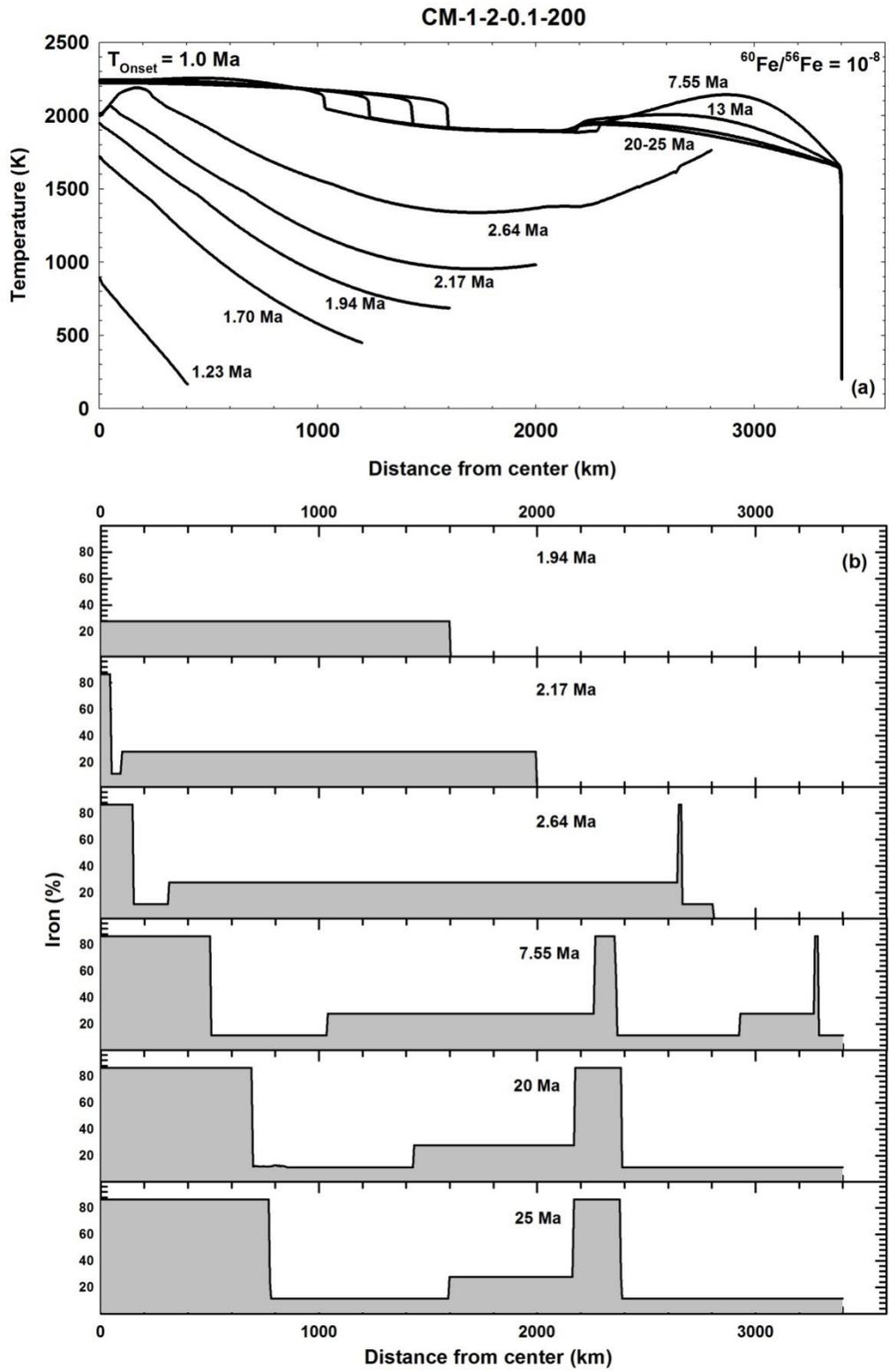

**Fig. 3.**

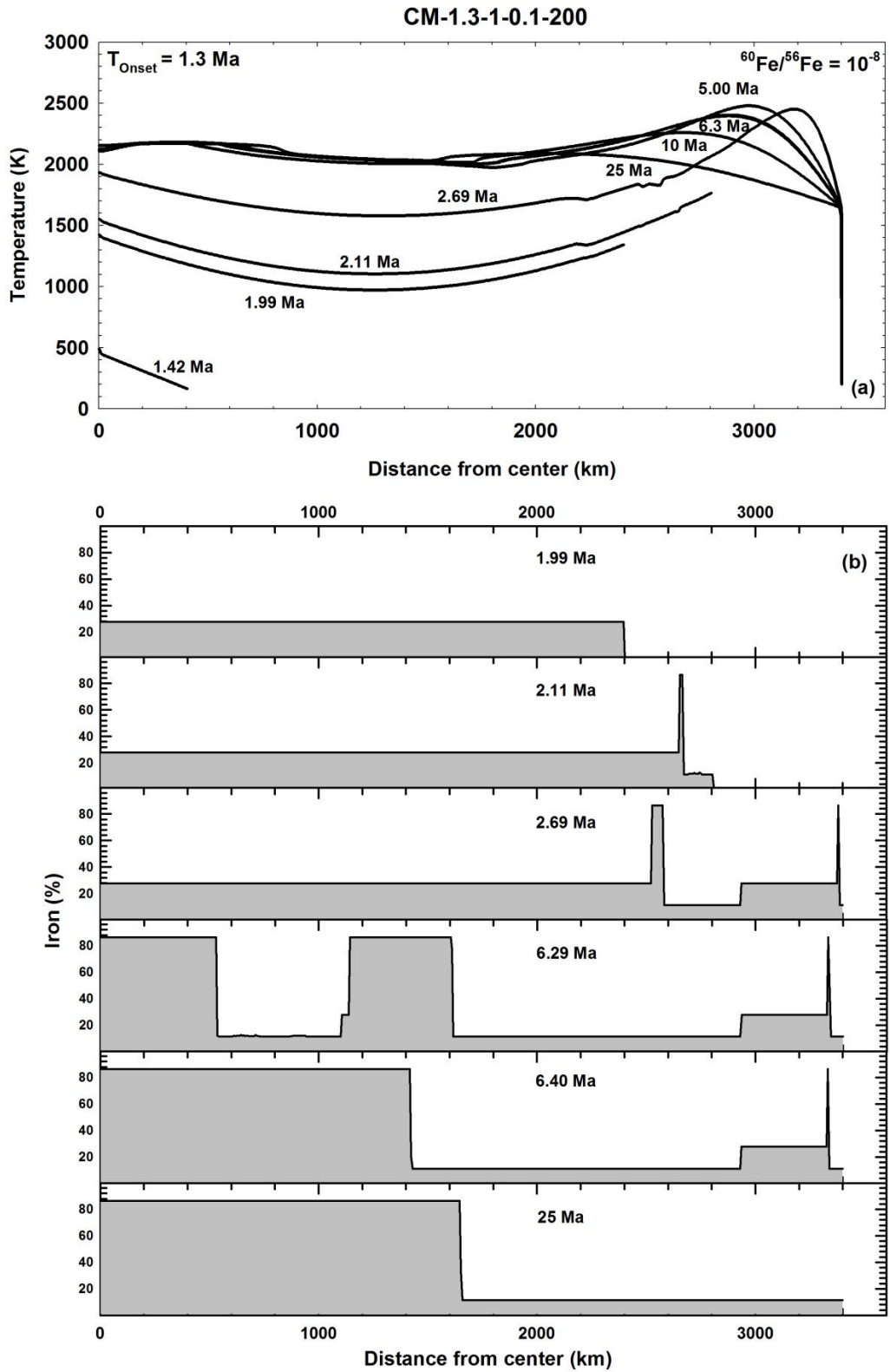

**Fig. 4.**

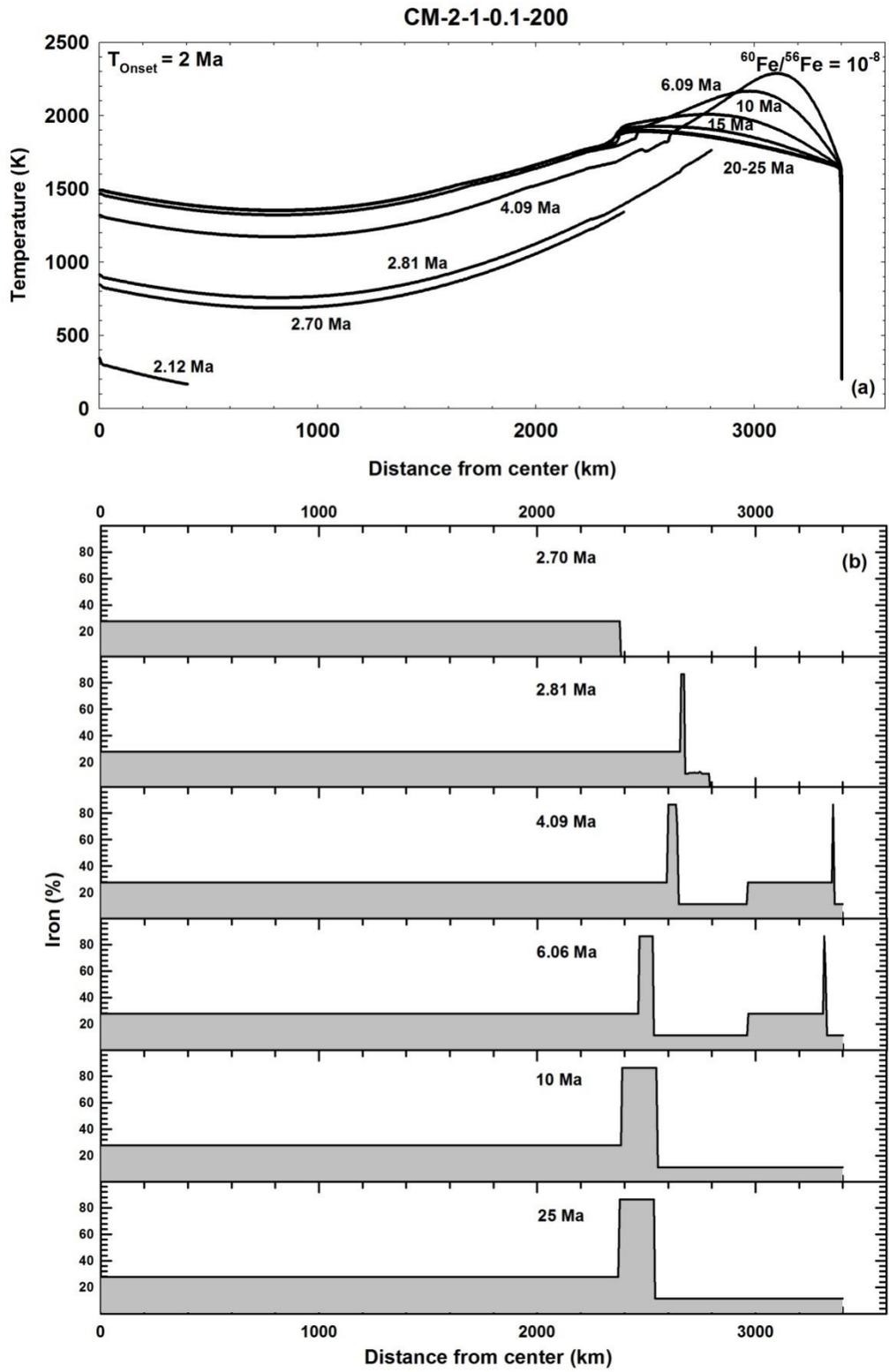

**Fig. 5.**

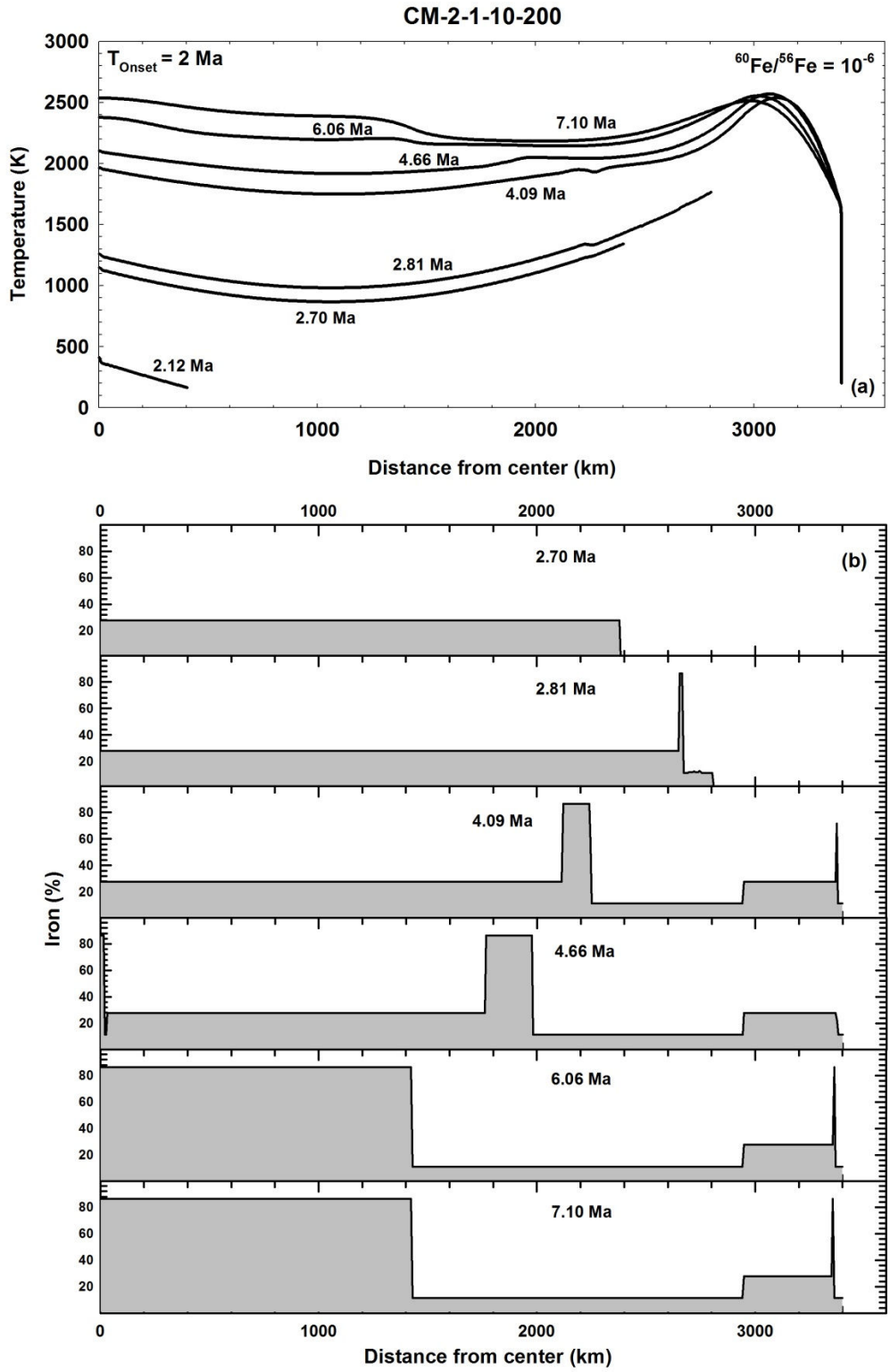

**Fig. 6.**

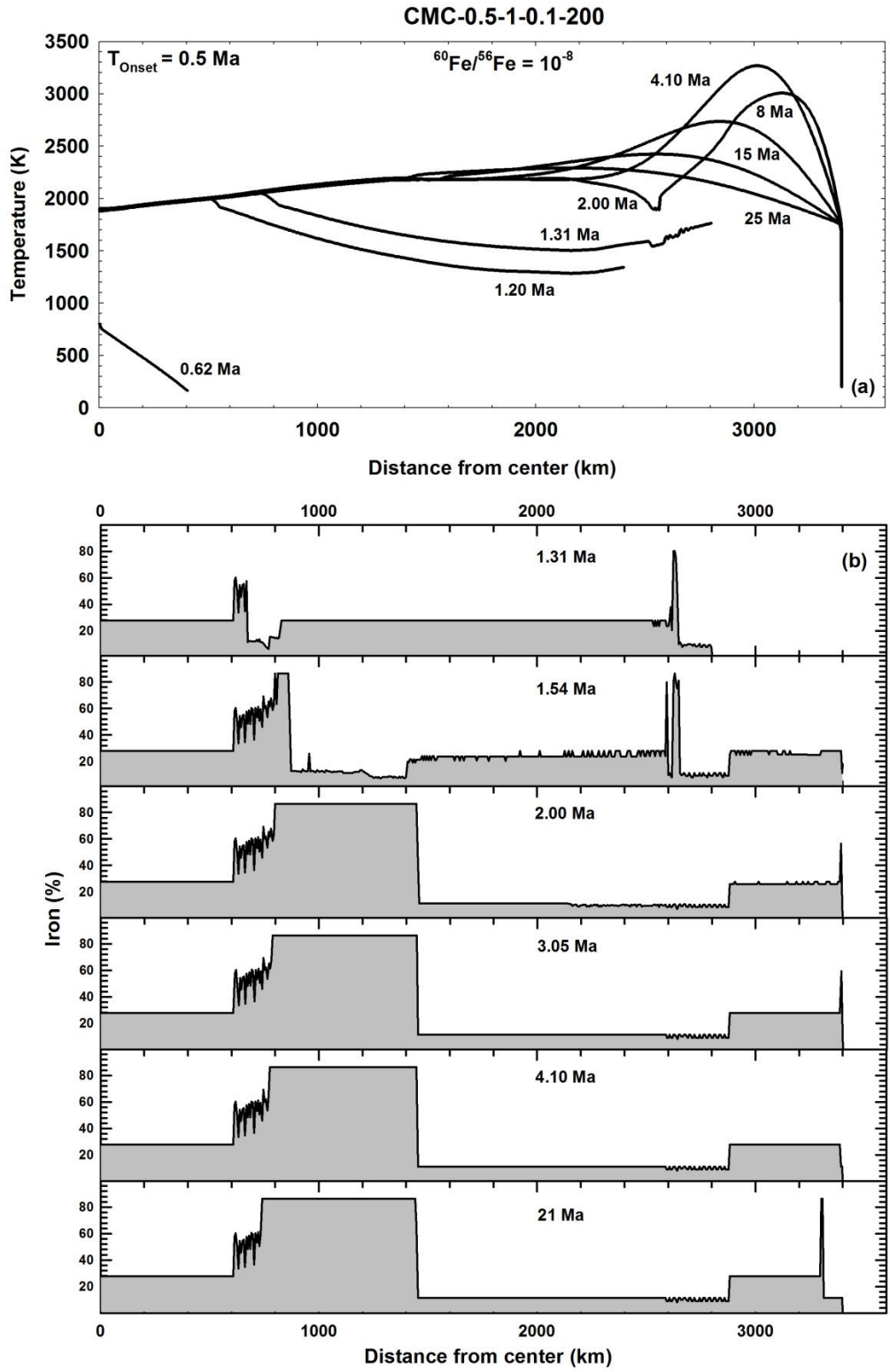

**Fig. 7.**

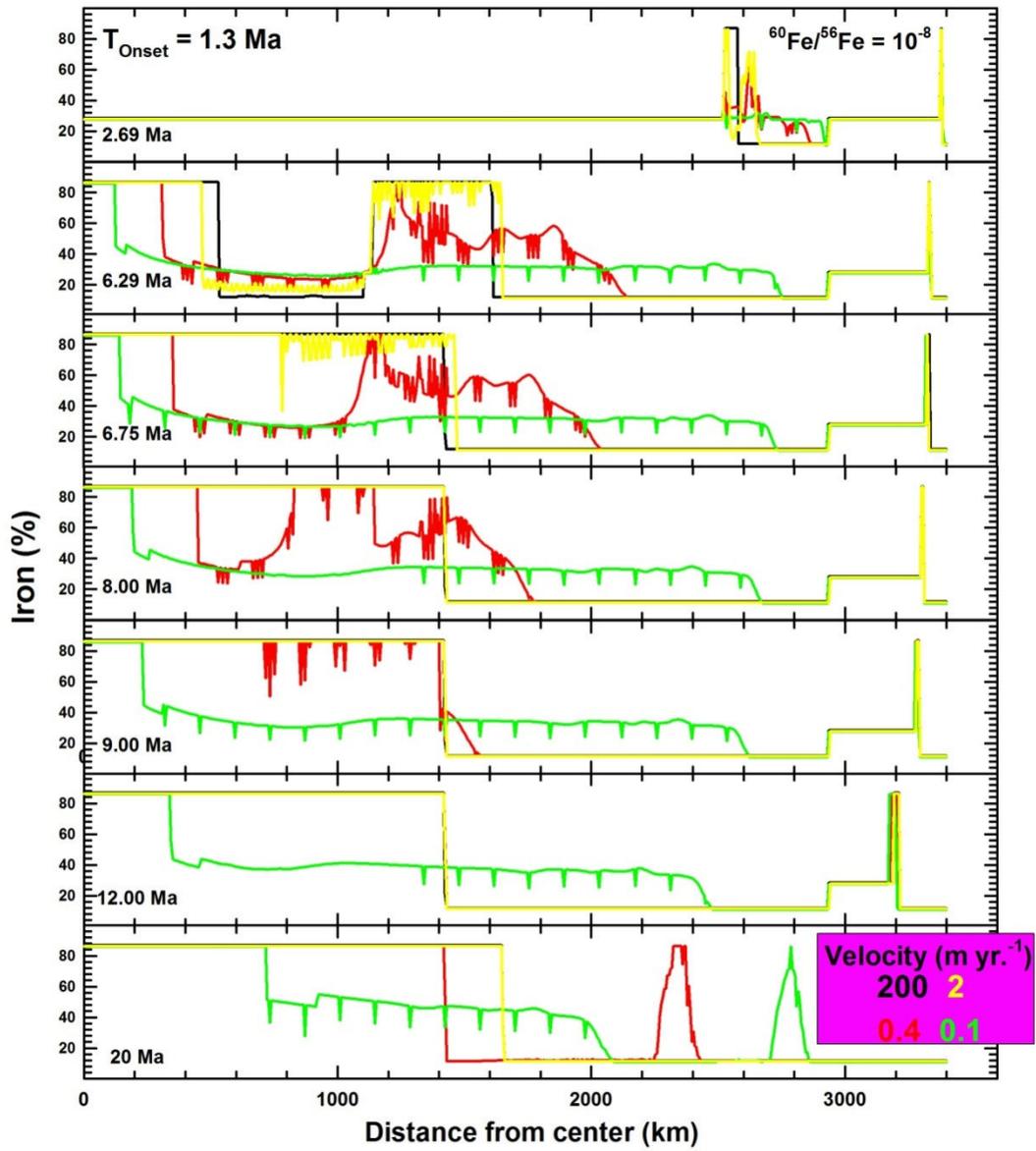

**Fig. 8.**